\documentclass[12pt]{article}
\usepackage{epsf,amsfonts,amssymb,epsfig,amsmath}
\renewcommand{\text}[1]{#1}

\newcommand{\be}{\begin{equation}}
\newcommand{\ee}{\end{equation}}
\newcommand{\ben}{\begin{displaymath}}
\newcommand{\een}{\end{displaymath}}
\newcommand{\bea}{\begin{eqnarray}}
\newcommand{\eea}{\end{eqnarray}}
\newcommand{\bean}{\begin{eqnarray*}}
\newcommand{\eean}{\end{eqnarray*}}
\newcommand{\nn}{\nonumber \\}
\newcommand{\ba}{\begin{array}}
\newcommand{\ea}{\end{array}}
\newcommand{\bi}{\begin{itemize}}
\newcommand{\ei}{\end{itemize}}

\newcommand{\reef}[1]{(\ref{#1})}

\def\l{\lambda}
\def\a{\alpha}

\def\b{\beta}

\def\d{\delta}

\def\e{\epsilon}
\def\s{\sigma}
\def\e{\epsilon}

\def\w{\wedge}
\def\r{\rho}
\def\sss{\sqrt{1-\lambda^2\rho^2}}
\def\ssst{\sqrt{\lambda^2\rho^2-1}}

\newcommand{\bbR}{{\mathbb{R}}}

\newcommand{\dd}{\mathrm{d}}

\newcommand{\vol}{\mbox{Vol}}

\begin{document}

\makeatletter
\renewcommand{\theequation}{\thesection.\arabic{equation}}
\@addtoreset{equation}{section} \makeatother

\baselineskip 18pt

\begin{titlepage}

\vfill

\begin{flushright}
Imperial/TP/2007/JG/01\\
\end{flushright}

\vfill

\begin{center}
   \baselineskip=16pt
   {\Large\bf $AdS$ Spacetimes From Wrapped $D3$-Branes}
   \vskip 2cm
      Jerome P. Gauntlett and Ois\'{\i}n A. P. Mac Conamhna
   \vskip .6cm
      \begin{small}
      \textit{Theoretical Physics Group, Blackett Laboratory, \\
        Imperial College London, London SW7 2AZ, U.K.}
        %E-mail: j.gauntlett, d.waldram@imperial.ac.uk}
        \end{small}\\*[.6cm]
      \begin{small}
      \textit{The Institute for Mathematical Sciences, \\
        Imperial College London, London SW7 2PE, U.K.}
        %E-mail: j.gauntlett, d.waldram@imperial.ac.uk}
        \end{small}
   \end{center}

\vfill

\begin{center}
\textbf{Abstract}
\end{center}
We derive a geometrical characterisation of a large class of 
$AdS_3$ and $AdS_2$ supersymmetric spacetimes in type IIB supergravity with non-vanishing
five-form flux using $G$-structures. These are obtained as special cases of a class of 
supersymmetric spacetimes with an $\bbR^{1,1}$ or $\bbR$ (time) factor 
that are associated with $D3$-branes wrapping calibrated 
2- or 3-cycles, respectively, in manifolds with $SU(2)$, $SU(3)$, $SU(4)$ and $G_2$ 
holonomy. We show how two explicit $AdS$ solutions, previously constructed in gauged
supergravity, satisfy our more general $G$-structure conditions. For
each explicit solution we also derive a special holonomy metric which, although singular,
has an appropriate calibrated cycle. 
After analytic continuation, some of the classes of $AdS$ spacetimes
give rise to known classes of BPS bubble solutions with $\bbR\times
SO(4)\times SO(4)$, $\bbR\times SO(4)\times U(1)$, and $\bbR\times SO(4)$ symmetry. These
have $1/2$, $1/4$ and $1/8$ supersymmetry, respectively. We present a
new class of 1/8 BPS geometries with $\bbR\times SU(2)$ symmetry,
obtained by analytic continuation of the class of $AdS$ spacetimes 
associated with $D3$-branes wrapped on associative three-cycles. 

\begin{quote}

\end{quote}

\vfill

\end{titlepage}
\setcounter{equation}{0}

%%%%%%%%%%%%%%%%%%%%%%%%%%%%%%%%%%%%%%%%%%%%%%%%%%%%%%%%%%%%%%%%%%%%%%%
%\tableofcontents
%%%%%%%%%%%%%%%%%%%%%%%%%%%%%

\section{Introduction}

A supersymmetric solution of D=10 or D=11 supergravity theory with
an anti-de-Sitter ($AdS$) factor is expected to be dual to a supersymmetric conformal field theory (SCFT). 
It is interesting to elucidate and study 
the geometrical structures underpinning such solutions for several reasons. 
For example, the results provide a good starting point for constructing explicit solutions.
More generally, a precise global characterisation of the relevant geometry is the first
step in attempting to obtain existence theorems.
Another application is to find geometrical analogues 
of general properties of classes of SCFTs as in 
\cite{Martelli:2005tp,Martelli:2006yb}. 
Finally, possibly after analytic continuation, the $AdS$
solutions can give rise to classes of ``bubble" solutions corresponding to
certain chiral primaries in SCFTs \cite{Lin:2004nb} or to 
solutions that describe supersymmetric defects, including Wilson lines, 
\cite{Yamaguchi:2006te,Lunin:2006xr,Gomis:2006cu,Lunin:2007ab,D'Hoker:2007xy,D'Hoker:2007fq,Gomis:2007fi}, 
all of which are interesting objects to study in the AdS/CFT correspondence.

It is by now well established that $G$-structure techniques \cite{Gauntlett:2002sc,Gauntlett:2002fz} 
are very useful in determining necessary 
and sufficient conditions for a class of geometries to give supersymmetric solutions. A key observation is that
the isotropy group of the Killing spinor(s) defines a canonical $G$-structure,
which can, for example, be characterised by certain bi-linears built from the spinors.
The Killing spinor equations then impose restrictions on the intrinsic torsion of the
$G$-structure and/or relate it to the flux.

To classify supersymmetric $AdS_{d+1}$ solutions one can start by assuming
that the metric is a general warped product of $AdS_{d+1}$ space, with its
maximally symmetric metric, and a Riemannian manifold ${N}$:
\be\label{adsa}
\dd s^2=\lambda^{-1}\dd s^2(AdS_{d+1})+ds^2({N})
\ee
where the warp factor $\lambda$ depends on the coordinates of ${N}$.
This ansatz is clearly invariant under the $SO(d,2)$ isometries of the $AdS$ 
space\footnote{For the $AdS_2$ case it is known that there are 
more general supersymmetric metrics with $SO(1,2)$ symmetry that arise
as the near horizon limit of rotating black holes \cite{Gauntlett:1998fz,Sinha:2006sh,Kunduri:2006uh}. 
It would be interesting to extend 
our analysis further to include these examples.}.
One also considers the most general $SO(d,2)$ invariant ansatz for the
matter fields (``fluxes") and then analyses the $G$-structures as just described. Some care is required in order to obtain a precise global statement about
the geometry: proceeding naively, say with bi-linears built from the Killing spinors, one might be working with $G$-structures
that are only locally defined\footnote{This is not a problem for the
specific application of constructing explicit solutions. 
This is because, in practise, one introduces local
coordinates, demands that the $G$-structure conditions are satisfied and then finally checks whether the local solution 
extends to a globally defined solution.}.

An alternative
strategy is to write the $AdS_{d+1}$ metric in Poincar\'e co-ordinates and
to consider the solution, locally, as a special case of a
supersymmetric solution with a $d$-dimensional Minkowski space factor. Thus if one
has an understanding of the geometry of spacetimes with Minkowski factors in terms of $G$-structures, then
one can extract out the geometry underlying the solutions with $AdS$
factors \cite{GMSW}. It turns out that it is in fact not necessary to
consider the most general Minkowski solutions: it was first shown in
\cite{GMSW} and then subsequently in \cite{Gauntlett:2006ux} that in many cases one can
consider classes of solutions with Minkowski factors that were called 
``wrapped-brane" solutions. The name arises because this class of Minkowski
solutions, by definition, preserves Killing spinors that satisfy the same
projections as for those of a probe brane wrapping a calibrated
cycle in a special holonomy manifold, or, equivalently, to a configuration
of intersecting branes. 
From a physical point of view this is in accord with our expectation
that the SCFTs dual to the $AdS$ solutions should live on such wrapped
or intersecting branes.

Thus a strategy to classify $AdS$ solutions is to first classify
wrapped-brane solutions, which is an interesting result in itself, and then
extract out the necessary and sufficient conditions for there to be
an $AdS$ solution. A nice feature of this approach is that it 
provides a neat global description of the relevant geometry arising in the
$AdS$ solution in
terms of the $G$-structure of the wrapped-brane solution \cite{Gauntlett:2006ux,Figueras:2007cn}. 
Typically, if we consider branes wrapping calibrated cycles in manifolds
with special holonomy $G$, the associated wrapped-brane solution will have a globally defined
$G$-structure but with non-trivial intrinsic torsion. 

The approach has now been used to classify $AdS$ solutions
associated with wrapped $M5$ branes \cite{Gauntlett:2006ux,Figueras:2007cn} and 
wrapped $M2$ branes \cite{MacConamhna:2006nb}. 
In this paper we will
focus on type IIB $AdS$ solutions associated with wrapped $D3$-branes.
In particular we will classify solutions that are a warped product of
a Minkowski spacetime with an internal space that are associated with
probe $D3$-branes wrapping associative 3-cycles, special lagrangian (SLAG) 3-cycles, and holomorphic 2-cycles in
Calabi-Yau ($CY$) 2, 3 and 4-folds. Furthermore, we assume that the only non-trivial flux is the self-dual five-form. 
We mentioned above that it is known that in many cases the procedure that we
will adopt leads to the most general classes of $AdS$ solutions of the type under consideration 
and we expect that to be the case here also. Our analysis will include
some cases that have already been studied before which will provide
some confirmation that this expectation is correct. It is certainly the case that in all cases the conditions that we 
derive are sufficient for having a supersymmetric $AdS$ solution.
Rather than insisting at all points on complete generality, our goal is to define a tractable framework
to explore the interplay of Anti-de Sitter and wrapped-brane geometry.

The cases that we shall consider
in this paper are summarised in Table~\ref{green}. We have listed the different 
types of calibrated cycles a probe $D3$-brane can wrap inside the given special holonomy manifold.
We have also listed the unwrapped world-volume of the $D3$-brane, with $\bbR^{1,1}$ representing two-dimensional Minkowski spacetime 
and $\bbR$ referring to a time direction,
along with the amount of supersymmetry on this space. The final column indicates the $R$-symmetry that arises
in the corresponding CFTs; this manifests itself as isometries in the classes of $AdS$ solutions that we shall derive.

\begin{table}
\begin{center}
\setlength{\tabcolsep}{0.45em}
\begin{tabular}{|c|c|c|c|c|}
\hline
wrapped brane & manifold & world-volume &
   susy& $R$-symmetry\\
\hline
Associative & $G_2$ & $\bbR$ & ${\cal N}=2$& $U(1)$\\
SLAG 3-cycle & $CY_3$ & $\bbR$ & ${\cal N}=4$&$SU(2)$\\
K\"ahler 2-cycle & $CY_4$ & $\bbR^{1,1}$ & ${\cal N}=(0,2)$&$U(1)$\\
K\"ahler 2-cycle & $CY_3$ & $\bbR^{1,1}$ & ${\cal N}=(2,2)$&$U(1)\times U(1)$\\
K\"ahler 2-cycle & $CY_2$& $\bbR^{1,1}$ & ${\cal N}=(4,4)$&$SO(4)\times U(1)$ \\
\hline
\end{tabular}
\end{center}
\caption{Wrapped $D3$-brane geometries and their supersymmetry}
\label{green}
\end{table}
In section 2 of this paper we will present and discuss the wrapped-brane geometries.
The wrapped-brane geometries corresponding to $D3$-branes wrapping associative 3-cycles 
and K\"ahler 2-cycles in $CY_4$ can be obtained as special cases of a more general classification of
type IIB geometries with five-form flux that was carried out
recently using $G$-structure techniques
in \cite{jan}. We will obtain the wrapped-brane geometries for all other 
cases by exploiting the fact that the geometries must 
admit multiple copies of these basic $G$-structures along with another assumption which we will explain in section 2.
This exactly parallels what was done in \cite{Gauntlett:2006ux} where it
was shown in that context that this indeed does give the most general 
wrapped-brane geometries. Since we strongly suspect that this is also true here, 
in the sequel we will refer to these geometries as wrapped-brane geometries.

In section 3, starting from the wrapped-brane geometries in section 2, we determine
the extra conditions that need to be imposed in order to obtain $AdS$ spacetimes.
The $AdS_2$ geometries for the associative and
SLAG 3-cycles that we derive are new. The $AdS_3$ geometries (or the
corresponding bubble solutions that we discuss in a moment) for the cases associated with 
$D3$-branes wrapping holomorphic 2-cycles in $CY_2$, $CY_3$ and $CY_4$ have all been considered before 
but the derivation from wrapped-brane geometries is new. It is satisfying that we find results in agreement
with \cite{Lin:2004nb,Kim:2005ez}.

It is sometimes possible to analytically continue $AdS$ solutions to obtain other classes of BPS solutions.
For example the classes of $AdS_3$ solutions arising from $D3$-branes wrapping holomorphic
2-cycles in $CY_2$, $CY_3$ and $CY_4$ give rise to known BPS ``bubble'' solutions with $\bbR\times SO(4)\times SO(4)$,
$\bbR\times SO(4)\times U(1)$ and $\bbR\times SO(4)$ symmetry, that
have $1/2$, $1/4$ and $1/8$ supersymmetry, respectively \cite{Lin:2004nb,Donos:2006iy,Donos:2006ms,Kim:2005ez} 
(for a unified discussion, differing from that given in this paper, see \cite{Chen:2007du}). 
Another class of BPS bubble geometries in $D=11$ supergravity can be found in \cite{Kim:2006qu} 
and further examples can be easily obtained from the results of \cite{Gauntlett:2006ux,MacConamhna:2006nb}
(see \cite{Kim:2007hv} for an alternative derivation of some of the cases considered in \cite{Gauntlett:2006ux,MacConamhna:2006nb}).
Here we will see that an analytic continuation of the $AdS_2$ geometries associated with 
$D3$-branes wrapped on associative 3-cycles leads to an interesting new class of 
1/8 BPS solutions with $\bbR\times SU(2)$ symmetry.

In section 4, we show that two explicit $AdS$ solutions that were
first constructed using gauged supergravity \cite{mn,oz,naka}, do indeed satisfy our conditions. 
This provides a very good check on our calculations, the details of which we mostly omit. Using these
results also allows us construct an ansatz for the wrapped-brane
geometries that could describe solutions that interpolate from
a special holonomy manifold to the explicit $AdS$ solutions (this
should be contrasted with the interpolating solutions that
correspond to a ``flow across dimensions" \cite{mn}). 
In particular, we show that
the ansatz includes singular special holonomy manifolds with
a calibrated cycle that provide a local model for
probe $D3$-branes wrapping the calibrated cycle. 

In section 5 we briefly conclude.

\section{Wrapped $D3$-Brane Geometries}
In this section we will discuss the wrapped-brane geometries associated with
$D3$-branes wrapping calibrated cycles in manifolds with special holonomy $G$.
By definition, these geometries are warped products of $d$-dimensional Minkowski spacetime
with a $10-d$ dimensional Riemannian manifold $M_{10-d}$
\be
\dd s^2=L^{-1}\dd s^2(\bbR^{1,d-1})+\dd s^2(M_{10-d})
\ee
where the Minkowski spacetime should be viewed as the unwrapped part of the $D3$-brane. Thus
for $D3$-branes wrapping two-cycles we have $d=2$ and for $D3$-branes wrapping
three-cycles we have $d=1$. Both the warp factor $L$ and the metric on $M_{10-d}$ are independent of the
coordinates of the Minkowski factor. Only the self-dual five-form flux is non-zero and it is also taken
to be invariant under the symmetries of the Minkowski factor. We will write
\be
F_5=\Theta+*_{10}\Theta.
\ee 
In all cases, $M_{10-d}$ will admit a globally defined $G$-structure with non-trivial torsion, related to the
five-form flux, and will preserve 1/2 as much supersymmetry as the type IIB solution with special holonomy $G$ and vanishing
five-form flux.
We also expect that all of the classes of solutions that we consider will admit Killing spinors that
satisfy the same projections\footnote{For a more precise discussion of this point 
we refer to section 2 of \cite{Gauntlett:2006ux}.} as those of a probe $D3$-brane wrapping a special holonomy manifold:

We first discuss the case associated with $D3$-branes wrapping associative three-cycles in manifolds with $G_2$ holonomy.
The relevant wrapped-brane geometry can simply be obtained by making a restriction on
the more general classification that appeared in \cite{jan}. From this case we then 
derive the wrapped-brane geometry associated with $D3$-branes wrapping SLAG three-cycles, making clear what we assume
in the derivation. Similarly, the case associated with $D3$-branes wrapping K\"ahler two-cycles in $CY_4$ can
be obtained from the more general classification that appeared in \cite{jan} and we then derive
the cases associated with $D3$-branes wrapping K\"ahler two-cycles in $CY_3$ and $CY_2$. A nice consistency check
is that the latter wrapped-brane geometry can also be derived from the case associated with $D3$-branes wrapping
SLAG three-cycles.

\subsection{Associative geometry}
We first define this geometry, and then discuss some of its features. For this case the metric and five-form flux can be written
\bea\label{28w}
\dd s^2&=&-L^{-1}\dd t^2+\dd s^2({\cal M}_7)+L\dd s^2(\bbR^2)\nn
\Theta&=&\dd(e^0\w\varphi).
\eea
where $\partial_t$ is Killing and $e^0=L^{-1/2}\dd t$. There is a globally defined (no-where vanishing) 
$G_2$ structure that is specified by 
an associative three-form $\varphi$ and co-associative four-form
$*_7\varphi$ defined on $\mathcal{M}_7$ and compatible with the metric $\dd s^2(\mathcal{M}_7)$. 
Furthermore, both $\varphi$ and the warp factor $L$ can depend on the coordinates of 
$\mathcal{M}_7$ and $\bbR^2$. 
We require that the intrinsic torsion of the $G_2$ structure is 
determined as follows: 
\bea\label{awbc} 
\dd(e^0\w\mbox{Vol}_7)&=&0,\nn
\vol[\bbR^2]\w\dd *_7\varphi&=&0,\nn 
\varphi\w\dd\varphi&=&0.
\eea
Finally we require that the Bianchi identity is satisfied,
\bea
\dd *_{10}\Theta=0.
\eea
With a hopefully obvious choice of frame, we take the ten-dimensional
orientation to be positive with respect to
\bea
\mbox{Vol}_{10}&=&e^0\w\mbox{Vol}[\mathcal{M}_{7}]\w
e^8\w e^9,\nn\mbox{Vol}[\mathcal{M}_7]&=&\frac{1}{7}\varphi\w *_7\varphi.
\eea

Having defined associative geometry we now begin to discuss its
features. Every solution of these equations will, by definition, 
admit two Killing spinors, which satisfy
the same orthonormal-frame projections as those of a probe $D3$-brane
wrapping an associative 3-cycle in a $G_2$ manifold. This is because
we have obtained the torsion conditions as a special case of those of
the more general
class of geometries called $G_2$ backgrounds in \cite{jan}. By the
construction of \cite{jan}, the associative geometries then admit two
Killing spinors, satisfying the appropriate algebraic constraints. In
order to get the wrapped-brane geometry of interest here we set 
$m=Y_{\pm}=0$ in section 6.1 of \cite{jan}. With some
work one can recast this restriction of the conditions of \cite{jan} in the 
more transparent way given above\footnote{Up to an irrelevant factor
of $-1/4$ in the definition of the five-form.}. For these geometries,
supersymmetry plus the Bianchi identity implies all the equations 
of motion of IIB supergravity are satisfied: this can be shown by studying the 
integrability conditions for type IIB \cite{Gauntlett:2005ww} (see also \cite{Gran:2005ct})
and generalising an argument presented in \cite{Gauntlett:2002fz}.

The wrapped-brane geometry should be able to describe back reacted $D3$-branes wrapping associative
3-cycles (or alternatively an appropriate configuration of intersecting $D3$-branes)
and it has several intuitive features. 
As we have noted before, the Killing time direction in \reef{28w} corresponds to the unwrapped part
of the $D3$-brane, while the two ``overall transverse'' directions, i.e.
transverse to the probe $D3$-brane world-volume and to the $G_2$ holonomy manifold, 
are visible as the $\bbR^2$ factor in \reef{28w}. Recall that the $\bbR^{1,2}\times X_7$ solution of type IIB supergravity where 
$X_7$ has $G_2$ holonomy preserves four supersymmetries and obviously has a globally defined
$G_2$. In the associative wrapped-brane geometry there is still a globally defined $G_2$ structure, 
but it now has non-trivial torsion which leads to the preservation of two supersymmetries.

As somewhat of an aside let us discuss how 
some of the conditions on the geometry can be understood in terms of generalised calibrations 
\cite{Gutowski:1999tu} (for further discussion 
and references see section 4 of \cite{Gauntlett:2006ux}). In particular, the expression
for the flux in \reef{28w} reveals that $\varphi$ is a generalised calibration:
it arises because the geometries can describe the back-reacted
geometry of $D3$-branes wrapping associative 3-cycles. 
Some of the other conditions in \reef{awbc} have a 
similar interpretation. The
first condition in \reef{awbc} states that in associative geometry, we can
also wrap a probe $D7$-brane over the
entire $G_2$ structure manifold while preserving supersymmetry. 
The second condition in \reef{awbc} seems to be related to
probe $D5$-branes that are calibrated by $*_7\varphi$
and one of the overall transverse directions.  
However, this condition is {\it not} equivalent to
\bea\label{notcal}
\dd (e^0\w*_7\varphi\w e^8)&=&0,\nn
\dd (e^0\w*_7\varphi\w e^9)&=&0
\eea 
as one might naively expect from such an interpretation 
(\reef{awbc} allows $*_7\varphi$ to have non-trivial 
dependence on the coordinates of $\bbR^2$, whereas \reef{notcal} does not).
Furthermore the last condition
in \reef{awbc} does not have any obvious interpretation in terms of generalised
calibrations. This example shows that one needs to use the intuition obtained from
generalised calibrations with care.

\subsection{SLAG-3 geometry}
In this subsection we define SLAG-3 geometry. The metric and the flux are given by
\bea\label{s31} \dd s^2&=&-L^{-1}\dd t^2+\dd s^2({\cal M}_6)+L\dd
s^2(\bbR^3),\nn \Theta&=&-\dd(e^0\w \mbox{Im}\Omega). 
\eea
where $\partial_t$ is Killing and $e^0=L^{-1/2}\dd t$.
We require the existence of a globally defined $SU(3)$ structure, given by an everywhere
non-vanishing (1,1) form $J_6$ and an everywhere
non-vanishing (3,0) form $\Omega_6$, defined on $\mathcal{M}_6$
and compatible with $\dd s^2({\cal M}_6)$. The $SU(3)$ structure and the warp
factor $L$ can depend on the coordinates of both $\mathcal{M}_6$ and $\bbR^3$.
The $SU(3)$ structure satisfies the following conditions on the intrinsic torsion:
\bea\label{swbc}
\dd J_6&=&0,\nn
\vol[\bbR^3]\w\dd(e^0\w\mbox{Re}\Omega_6)&=&0,\nn 
\mbox{Im}\Omega_6\w\dd\mbox{Im}\Omega_6&=&0. \eea 
We also demand that $\dd *_{10}\Theta=0$.

These geometries admit four Killing spinors
(half-maximal for an $SU(3)$ structure), satisfying the appropriate
algebraic constraints. This is because we have derived the above class of
geometries by requiring {\it two} independent associative structures,
each of which implies the existence of two Killing spinors. 
If we write \bea \varphi&=&\pm J_6\wedge
e^7-\mbox{Im}\Omega_6, \nn *_7\varphi&=&\tfrac{1}{2}J_6\wedge
J_6\pm \mbox{Re}\Omega_6\wedge e^7, \eea and set
$e^7=L^{1/2}\dd y$, after substituting into
\eqref{28w}, \reef{awbc}, we get torsion conditions given in the
definition. The requirement that we have two independent associative structures is exactly 
what is wanted for the geometries to preserve Killing spinors satisfying the same
algebraic constraints as probe $D3$-branes wrapping SLAG 3-cycles in the sense of \cite{Gauntlett:2006ux}.
Note that our assumption that $e^7=L^{1/2}\dd y$ implies that there is an $\bbR^3$ factor in \reef{s31}
corresponding to the three overall transverse directions of a probe $D3$-brane wrapping a SLAG three-cycle.
Based on the results of \cite{Gauntlett:2006ux} we strongly suspect that this assumption
is actually implied by demanding 
that the class of geometries admit Killing spinors satisfying
the same projections as wrapped probe branes. In this paper we will be content 
to just assume this condition, and hence it effectively becomes part of our definition of
a SLAG-3 geometry. Similar comments will apply to other geometries that we discuss below.

As in the associative case some of these conditions can be
interpreted, if somewhat imprecisely,  in terms
of generalised calibrations. The expression for the flux says that
-$\mbox{Im}\Omega_6$ is a generalised calibration corresponding to
the fact that the geometries can describe the back-reacted geometry of 
$D3$-branes wrapping SLAG 3-cycles. 
The first condition in \reef{swbc}, which can be written $d(e^0\w J_6\w e^i)=0$, where $e^i$ is a frame direction
in any of the three overall transverse directions, corresponds to probe $D3$-branes wrapping a 
cycle calibrated by $J_6$ and any one of the overall transverse directions. 
The second condition in \reef{swbc} corresponds to probe $D5$-branes that
are calibrated by $\mbox{Re}\Omega$ and any two of the three overall transverse directions, 
but as in the associative case, only imprecisely. 
The last condition in \reef{swbc} doesn't have any obvious generalised calibration interpretation.

\subsection{K\"ahler-2 in $CY_4$ geometry}
The metric and flux for this geometry take the form
\bea \label{mcy41}
\dd s^2&=&L^{-1}\dd s^2(\bbR^{1,1})+\dd s^2({\cal M}_8),\nn
\Theta&=&\vol[\bbR^{1,1}]\w\dd (L^{-1}J_8).
\eea
We now require the existence of a globally-defined $SU(4)$
structure, specified by everywhere-non-zero forms $J_8$, $\Omega_8$
on $\mathcal{M}_8$ and compatible with the metric $\dd s^2({\cal M}_8)$.
The intrinsic torsion conditions can be written
\bea \label{mcy42}
\dd(L^{-1}J_8\w J_8\w J_8)&=&0,\nn
%=\dd* J&=&0,\nn
\dd(L^{-1}\Omega_8)&=&0,
%\dd*\Omega&=&0.
\eea
and we also demand that $\dd *_{10}\Theta=0$. Positive orientation is
defined with respect to
\bea
\mbox{Vol}_{10}&=&e^0\w
e^1\w\mbox{Vol}[\mathcal{M}_8],\nn
\mbox{Vol}[\mathcal{M}_8]&=&\frac{1}{4!}J_8\ J_8\w J_8\w J_8.
\eea

We have obtained the torsion conditions from the $SU(4)\ltimes \bbR^8$
case\footnote{A potentially confusing point is that the
  isotropy group of the Killing spinors in the geometry of \cite{jan} is
  $SU(4)\ltimes\bbR^8$. However, the assumption that we have a warped product of $\bbR^{1,1}$ 
with an eight-dimensional manifold reduces this to an $SU(4)$ structure.}
 of \cite{jan}; they were also derived in \cite{jn}. It follows from the construction of \cite{jan} that these
geometries admit two Killing spinors, half-maximal for an $SU(4)$ structure. In an orthonormal frame these
satisfy the same algebraic projections as for probe $D3$-branes wrapping a K\"ahler two-cycle
in a $CY_4$. The Killing spinors are pure.
Note that for this case there are no overall transverse directions. 

The expression for the flux says that
$J_8$ is a generalised calibration corresponding to
the fact that the geometries can describe back-reacted $D3$-branes
wrapping holomorphic 2-cycles. 
The first condition in \reef{mcy42} corresponds to probe $D7$-branes that
wrap a 6-cycle calibrated by $J_8^3/3!$,
while the second condition corresponds to probe $D5$-branes wrapping a 
four-cycle calibrated by the real or imaginary part of $\Omega_8$.

\subsection{K\"ahler-2 in $CY_3$ geometry}
We next consider the wrapped-brane geometries 
associated with $D3$-branes wrapping holomorphic two-cycles in a Calabi-Yau three-fold. 
The metric and the flux are given by
\bea\label{mcy31}
\dd s^2&=&L^{-1}\dd s^2(\bbR^{1,1})+\dd s^2({\cal M}_6)+L\dd s^2(\bbR^{2}),\nn
\Theta&=&\mbox{Vol}[\bbR^{1,1}]\wedge \dd(L^{-1} J_6).
\eea
We demand the existence of a globally-defined $SU(3)$ structure, with everywhere
non-vanishing structure forms $J_6$, $\Omega_6$ defined on $\mathcal{M}_6$ and
compatible with $\dd s^2({\cal M}_6)$. 
The warp factor $L$ and $J_6$, $\Omega_6$ 
depend on the coordinates of both $\mathcal{M}_6$ and $\bbR^2$.
We require the following torsion conditions
\bea\label{mcy32}
\vol[\bbR^2]\w \dd(J_6\w J_6)&=&0,\nn
%\dd(L^{-1}J^3)&=&0,\nn
\dd(L^{-1/2}\Omega_6)&=&0,
\eea
and also demand that $\dd *_{10}\Theta=0$.

The torsion conditions may be derived by assuming a pair of $SU(4)$
structures satisfying the K\"ahler-2 in $CY_4$ conditions of the last subsection. To see this, write
\bea J_8&=&J_6\pm e^7\wedge e^8,\nn
\Omega_8&=&\Omega_6\wedge (e^7\pm ie^8),
\eea
together with $e^7=L^{1/2}\dd x^7$, $e^8=L^{1/2}\dd x^8$.
Substituting both these structures into \reef{mcy41}, \reef{mcy42}
produces \reef{mcy31} and \reef{mcy32}.

The expression for the flux says that
$J_6$ is a generalised calibration corresponding to
the fact that the geometries can describe back-reacted $D3$-branes
wrapping holomorphic 2-cycles. 
The first condition in \reef{mcy32} corresponds to probe $D7$-branes that
wrap a 4-cycle calibrated by $J_6^2/2$ and two of the overall transverse directions (the $\bbR^2$ factor). 
The second condition corresponds to probe $D5$-branes wrapping a 
3-cycle calibrated by the real or imaginary part of $\Omega_6$ and one of the two overall transverse directions.
To see this we note that we can rewrite the condition as
$\dd (e^0\w e^1\w \Omega_6 \w e^i)=0$, for arbitrary $e^i$ on the transverse space.
Furthermore, we note that the last condition implies that 
$\dd(L^{-1}\vol[\mathcal{M}_6])=0$ which corresponds to probe $D7$-branes wrapping the whole
$CY_3$.

\subsection{K\"ahler-2 in $CY_2$ geometry}
Finally we consider the wrapped-brane geometry corresponding to
$D3$-branes wrapping holomorphic two-cycles in a Calabi-Yau two-fold. 
The metric and flux are given by 
\bea\label{su21}
\dd s^2&=&L^{-1}\dd
s^2(\bbR^{1,1})+\dd s^2({\cal M}_4)+L\dd s^2(\bbR^4),\nn
\Theta&=&\mbox{Vol}[\bbR^{1,1}]\wedge \dd(L^{-1} J_4). \eea
There is a globally defined $SU(2)$ structure, with
nowhere-vanishing structure forms $J_4$, $\Omega_4$ defined on $\mathcal{M}_4$ and
compatible with the metric $\dd s^2({\cal M}_4)$.
The warp factor $L$ and $J_4$, $\Omega_4$ 
depend on the coordinates of both $\mathcal{M}_4$ and $\bbR^4$.
We require the torsion conditions
\bea\label{su22} \vol[\bbR^4]\w\dd(LJ_4)
&=&0,\nn
 \dd\Omega_4&=&0, \eea
and that $\dd *_{10}\Theta=0$.

These conditions imply the existence of eight Killing spinors,
half-maximal supersymmetry for an $SU(2)$ structure. To see this,
observe that they imply the existence of two $SU(3)$ structures
satisfying the constraints of the previous subsection. The $SU(3)$
structures are  
\bea 
J_6&=&J_4\pm e^5\wedge e^6,\nn
\Omega_6&=&\Omega_4\wedge(e^5\pm ie^6), 
\eea
with $e^5=L^{1/2}\dd x^5$ and $e^6=L^{1/2}\dd x^6$. Requiring that
both these structures satisfy the torsion conditions of the previous
subsection, we get \reef{su21} and \eqref{su22}.

The expression for the flux says that
$J_4$ is a generalised calibration corresponding to
the fact that the geometries describe back-reacted $D3$-branes
wrapping holomorphic 2-cycles. 
The first condition in \reef{su22} corresponds to probe $D7$-branes that
wrap a 2-cycle calibrated by $J_4$ and four overall transverse directions. 
The second condition corresponds to probe $D5$-branes wrapping a 
2-cycle calibrated by the real or imaginary part of $\Omega_4$ 
and two of the four overall transverse directions: 
to see this we note that the condition can be equivalently 
$\dd (e^0\w e^1\w\Omega_4\w
  e^i\w e^j)=0$
for arbitrary $e^i$, $e^j$ on the overall transverse space.

We can also obtain these 
conditions starting from the SLAG-3 case, which provides a nice consistency
check.
In particular, if we decompose the $SU(3)$ structure $J_6,\Omega_6$ as
\bea J_6&=&J^{(3)}\pm e^5\wedge e^6,\nn \Omega_6&=&(J^{(2)}+iJ^{(1)})\w(e^5\pm ie^6),
\eea
set $e^5=L^{1/2}\dd x^5$, $e^6=L^{1/2}\dd x^6$,  and substitute into the SLAG-3 geometry conditions
\reef{s31}, \reef{swbc} we recover \reef{su21} and \reef{su22}, provided that we identify
$J^{(1)}$ and $J^{(3)}+iJ^{(2)}$ with $J_4$ and $\Omega_4$, respectively.

Finally, we point out that we strongly suspect that
\reef{su21} and \reef{su22} do not constitute the most general 
wrapped-brane spacetimes in this class. In particular,
by analogy with a similar case that was studied in \cite{MacConamhna:2006nb}, we expect that
the derivation that we have used misses the 
possibility that the overall transverse space, $\bbR^4$ in  
\reef{su21} and \reef{su22} can be replaced with an arbitrary
$CY_2$ metric. It would be interesting to verify this. 
On the other hand we suspect that in
order to obtain an $AdS_3$ limit, as we do in the next section, it is necessary to take the $CY_2$
metric to be flat $\bbR^4$; again, this would be interesting to verify
explicitly. 

\section{AdS and Bubble Geometries}

In the previous section we have defined classes of wrapped-brane 
solutions of type IIB supergravity. In all cases the metric has the form
\bea
\dd s^2=L^{-1}\dd s^2(\bbR^{1,d-1})+\dd s^2({M}_{10-d})
\eea
For all cases, except the K\"ahler-2 in $CY_4$ case, there are at least two overall transverse
directions and we can write
\be
\dd s^2({M}_{10-d})=\dd s^2({\cal M}_G)+L\left[\dd z^2+z^2\dd s^2(S^{q})\right]
\ee
where $\dd s^2(S^{q})$ is the round metric on a $q$-sphere and the 
cases $q=1,2$ and $3$ appear. 
  
In this section we determine the extra conditions that need to be placed
on these geometries in order to extract an $AdS$ solution of the form
\bea\label{adsa2}
\dd s^2&=&\lambda^{-1}\dd s^2(AdS_{d+1})+\dd s^2(N_{9-d})\nn
&=&\lambda^{-1}\left[e^{-2r}\dd s^2(\bbR^{1,d-1})+\dd r^2\right]+\dd s^2(N_{9-d})
\eea
where in the second line we have written the unit radius 
$AdS$ space in Poincar\'e coordinates. We require that $\partial_r$ is a Killing
vector for $\dd s^2(N_{9-d})$.
Clearly to obtain this metric from any of the wrapped-brane geometries we
must insist that the warp factor takes the form
\be
L=e^{2r}\lambda.
\ee
For the K\"ahler-2 in $CY_4$ case, we must demand that $\dd s^2(M_8)$ is a cone in order
to extract out the $AdS$ radial direction, as we shall explain later (this
case is very analogous to the case of Sasaki-Einstein manifolds). For the
other cases, following \cite{Gauntlett:2006ux}, the unit radial one-form can be written
 \be\lambda^{-1/2}\dd r = \sin\theta\,\hat{u} + \cos\theta\,\hat{v},
\ee
where $\hat{u}$ is a unit one-form in $\mathcal{M}_G$, and
$\hat{v}$ is a unit one-form in the overall transverse space.
We will make the assumption that $\hat{v}$ is given by
 \begin{equation}
\label{t-ansatz}
   \hat{v} = L^{1/2}\dd z,
\end{equation}
and so lies along the radial direction $\dd z$ of the conformally flat
overall transverse space. In addition we will assume that the rotation
angle $\theta$ must be independent of the $AdS$ radial coordinate.
It seems likely that these assumptions can be relaxed (see \cite{Gauntlett:2006ux})
but we shall not investigate this issue further here.

We next introduce the orthogonal combination                    
\begin{equation}
   \hat{\rho} = \cos\theta\,\hat{u} - \sin\theta\,\hat{v} ,
\end{equation}
It is also convenient to introduce a new coordinate $\rho$ via
\be
\cos\theta=\lambda\rho,\qquad \sin\theta=\sqrt{1-\lambda^2\rho^2}.
\ee                                               
Then using the fact that $\dd z$ is closed, and $\theta$ is independent of $r$,
we find
\bea
\hat\rho&=&\frac{\lambda^{1/2}}{\sqrt{1-\lambda^2\rho^2}}\dd\rho,\nn
\hat u&=&\l^{-1/2}\sss\dd r+\frac{\l^{3/2}\rho}{\sss}\dd\rho.
\eea
In addition we also have
\bea
z&=&-e^{-r}\rho
\eea

We now write 
\bea
\dd s^2({M}_{10-d})&=&\dd s^2({\cal M}_G)+L\left[\dd z^2+z^2\dd s^2(S^{q})\right]\nn
&=&\dd s^2({\cal M}_{G'})+(\hat u)^2+(\hat v)^2+ Lz^2\dd s^2(S^{q})
\eea
where the $G'$-structure on ${\cal M}_{G'}$
is a reduction of the $G$-structure on ${\cal M}_G$ defined
by picking out the particular one-form $\hat u$. Given the above
formulae, we thus conclude that 
\be\label{N-metric}
\dd s^2(N_{9-d})=\dd s^2({\cal M}_{G'})+(\hat \rho)^2+\lambda\rho^2\dd s^2(S^{q})
\ee
Given the supersymmetry conditions on the original space
$M_{10-d}$ it is then straightforward to take~\eqref{adsa2}
with $\dd s^2(N_{9-d})$ given
by~\eqref{N-metric}, demand that the flux has no components along the
$AdS$ radial direction, and hence derive the supersymmetry conditions for an
$AdS_{d+2}$ geometry in terms of the $G'$-structure. We shall present the results of
these calculations, which can be technically involved, in the following sub-sections.
It is worth emphasising that, unlike the $G$-structure, this $G'$-structure is,
in general, only locally defined, since there can be points 
where $\sin\theta=0$ and hence the vector $\hat u$ is ill-defined.

The discussion thus far has been for the generic case where $\dd r$ 
lies partly in $\mathcal{M}_G$ and partly in the overall transverse
space. It is not hard to see that it is inconsistent for 
$\dd r$ to lie entirely in $\mathcal{M}_G$. 
One can also consider the possibility that
$\dd r$ lies entirely in the overall transverse space.
For cases where the torsion conditions imply a constraint
of the form
\begin{equation}
\label{vol-cond}
   \dd ( L^m \vol[\mathcal{M}_G] ) = 0 ,
\end{equation}
for some $m\ne 0$, it is also inconsistent. This leaves this possibility open for
two classes, SLAG 3-cycle and the 2-cycle in $CY_2$, and we shall comment on them
below. We now present the general $AdS$ limits, as described above, for each wrapped-brane geometry.

\subsection{$AdS_2$ from associative}
Writing $\dd s^2({\cal M}_7)=\dd s^2({\cal M}_6)+(\hat u)^2$, after the frame rotation we find that
the metric and flux are given by
\bea
\dd s^2&=&\frac{1}{\l}\left[\dd
  s^2(AdS_2)+\frac{\lambda^2}{1-\l^2\r^2}\dd\r^2+\l^2\r^2\dd
    s^2(S^1)\right]+\dd s^2(\mathcal{M}_6),\nn
\Theta&=&\mbox{Vol}[AdS_2]\w[-\dd(\l^{-1}\sss
J_6)+\l^{-1/2}\mbox{Im}\Omega_6-\l^{1/2}\r J_6\w\hat{\r}].
\eea
Here $\mathcal{M}_6$ has an $SU(3)$ structure $J_6,\Omega_6$.  
 We find that the $S^1$ direction is Killing, 
leaving both the $SU(3)$ structure and the warp factor $\l$ invariant. 
In addition the $SU(3)$ structure must satisfy
the conditions
\bea
\dd(\l^{-1/2}\mbox{Im}\Omega_6-\l^{1/2}\r J_6\w\hat{\rho})&=&0,\nn
\dd \left(\frac{1}{2}J_6\w J_6+\frac{1}{\l\rho}\mbox{Re}\Omega_6\w\hat{\r}\right) &=&0.
\eea
The result of a long calculation gives 
\bea
{}*_{10}\Theta=
\mbox{Vol}[S^1]\w\dd\left(\lambda^{-1/2}\sss\mbox{Re}\Omega_6\right).
\eea 
which shows that the Bianchi identity is satisfied.

These geometries are dual to SCQM with two supersymmetries.
The $U(1)$ isometry corresponds to the $U(1)$ $R$-symmetry of the dual
theory. 
An example of this geometry was constructed in \cite{oz} (see also \cite{naka}) and
we shall verify this directly in the next section.

It is interesting to notice that we can analytically continue these solutions
to obtain a new class of 1/8 BPS solutions with $\bbR\times SU(2)$ symmetry.
In particular, we take $\dd s^2(AdS_2)\to -\dd s^2(S^2)$,
$\mbox{Vol}[AdS_2]\to i\mbox{Vol}[S^2]$ and $\lambda\to -\lambda$ to get
\bea
\dd s^2&=&\frac{1}{\l}\left[\dd
  s^2(S^2)+\frac{\l^2}{\l^2\r^2-1}\dd\r^2-\l^2\r^2\dd
    T^2\right]+\dd s^2(\mathcal{M}_6),\nn
\Theta&=&\mbox{Vol}[S^2]\w\left[-\dd\Big(\l^{-1}\ssst
J_6\Big)+\l^{-1/2}\mbox{Im}\Omega_6+\frac{1}{\ssst}\l\r J_6\w \dd\r\right],\nn
&&
\eea
where the time coordinate $T$ was originally a coordinate on the $S^1$.
The torsion conditions are now
\bea
\dd\left(\l^{-1/2}\mbox{Im}\Omega_6+\frac{\l\r}{\ssst} J_6\w\dd\r \right)&=&0,\nn
\dd \left(\frac{1}{2}J_6\w J_6
+\frac{1}{\l^{1/2}\rho\ssst}\mbox{Re}\Omega_6 \w\dd\rho\right) &=&0.
\eea
It would be interesting to study this class of solutions further.

\subsection{$AdS_2$ from SLAG-3}

\paragraph{AdS radial direction from overall transverse space}
For this case it is possible for the radial direction to come from the
overall transverse space. One finds that $\lambda$ must be a constant, which we take to be 1,
and that the solution is simply the well known $AdS_2\times S^2\times CY_3$ solution:
\bea
\dd s^2&=&\dd s^2(AdS_2)+\dd s^2(S^2)+\dd s^2(\mathcal{M}_6),\nn
\Theta&=&\mbox{Vol}[AdS_2]\w \mbox{Im}\Omega_6.
\eea

\paragraph{AdS radial direction from frame rotation}
Alternatively we can have $\dd r$ point partially in the overall transverse direction
and partially in the direction of ${\cal M}_6$. Writing 
$\dd s^2({\cal M}_6)=\dd s^2({\cal M}_4)+(e^5)^2+(\hat u)^2$, after the frame rotation we find that
the metric and flux are
\bea
\dd s^2&=&\frac{1}{\l}\left[\dd
  s^2(AdS_2)+\frac{\l^2}{1-\l^2\r^2}\dd\r^2+\l^2\r^2\dd
    s^2(S^2)\right]+\dd s^2(\mathcal{M}_4)+e^5\otimes e^5,\nn
\Theta&=&\mbox{Vol}[AdS_2]\w[\dd(\l^{-1}\sss
J^3)+\l^{1/2}\r J^3\w\hat{\r}+\l^{-1/2}J^2\w e^5].
\eea
Here $\mathcal{M}_4$ has an $SU(2)$ structure $J^i$, $i=1,2,3$, with
$J^iJ^j=-\d^{ij}+\e^{ijk}J^k$. 
%We can write $J^1=J_4$ and $J^3+iJ^2=\Omega_4$. 
We find that the $S^2$ directions are Killing, preserve the $SU(2)$  
structure and that in addition
\bea
\dd(\l^{-1/2}\sss e^5)&=&0,\nn
\dd J^1&=&-\l\rho\dd\log\left(\frac{\l^2}{1-\l^2\r^2}\right)\w
e^5\w\hat{\r},\nn
\dd(\l^{1/2}\r J^3\w e^5)&=&\dd(\l^{-1/2}J^2\w\hat{\r}),\nn
\dd(\l^{1/2}\r J^3\w \hat{\r})&=&-\dd(\l^{-1/2}J^2\w e^5).
\eea
Taking the Hodge dual of the $\Theta$, a long calculation leads to 
\bea
{}*_{10}\Theta=-\mbox{Vol}[S^2]\w\Big[\dd\left(\r\sss
J^2\right)-\l^{-1/2}J^2\w\hat{\r}+\l^{1/2}\r J^3\w e^5\Big]
\eea
and we see that the Bianchi identity is again implied by the torsion conditions. 

These geometries are dual to SCQM with four supersymmetries.
The $SU(2)$ isometry of these manifolds is to be identified with the $R$-symmetry of the dual
quantum mechanics. We are unaware of any explicit solutions in this class.

At first glance it would seem that we could obtain a new class
of $AdS_2$ geometries by making the analytic continuation $\lambda\to-\lambda$,
$ds^2(S^2)\leftrightarrow ds^2(AdS_2)$. However, in the new solution if we
make a further redefinition $\lambda\to1/\lambda\rho^2$, along with
$J^2\leftrightarrow J^3$, we find that the solution is exactly the same as that above.

\subsection{$AdS_3$ from K\"{a}hler-2 in $CY_3$}
Writing $\dd s^2({\cal M}_6)=\dd s^2({\cal M}_4)+(e^5)^2+(\hat u)^2$, after the frame rotation 
we find that the metric and flux are
\bea
\dd s^2&=&\frac{1}{\l}\left[\dd
  s^2(AdS_3)+\frac{\l^2}{1-\l^2\r^2}\dd\r^2+\l^2\r^2\dd
    s^2(S^1)\right]+\dd s^2(\mathcal{M}_4)+e^5\otimes e^5,\nn
\Theta&=&\mbox{Vol}[AdS_3]\w[\dd(\l^{-3/2}\sss
e^5)-2\l^{-1}J_4-2\r e^5\w\hat{\r}].
\eea
$\mathcal{M}_4$ has an $SU(2)$ structure $J_4$, $\Omega_4$. We find that
the $S^1$ direction preserves the $SU(2)$ 
structure and is Killing. In addition we must have
\bea
\dd(\l^{-1}J_4+\r e^5\w\hat{\r})&=&0,\nn
\dd(\l^{-1}\sss\Omega_4)&=&i\l^{-1/2}\Omega_4\w
e^5-\l^{1/2}\r\Omega_4\w\hat{\rho}.
\eea
and that $\dd\lambda$ has no component in the $e^5$ direction.
These conditions imply that we can introduce a coordinate $\psi$
such that $e^5=A(\dd \psi+B)$ with $\partial_\psi$ a Killing vector and
$A=\l^{-1/2}\sss$. The Killing vector $\partial_{\psi}$ preserves $J_4$, though $\Omega_4$ has non-zero charge
under it. In these coordinates  The Hodge dual of $\Theta$ is 
\bea
{}*_{10}\Theta=-\mbox{Vol}[S^1]\w\dd\Big(\l^{-1/2}\sss J_4\w e^5\Big).
\eea
and we see that the Bianchi identity is again implied by the torsion conditions. 

These geometries are dual to two-dimensional SCFTs with $(2,2)$ supersymmetry.
These have a $U(1)\times U(1)$ $R$-symmetry which is dual to
the two $U(1)$ Killing vectors associated with the $S^1$ and $\partial_\psi$.
An example of this geometry can be found in \cite{mn} (see also
\cite{naka}): we will verify that this is indeed a solution of the
torsion conditions in the next section.

After analytic continuation we obtain BPS bubbles with 1/4 supersymmetry
and $\bbR\times SO(4)\times U(1)$ symmetry. These
conditions should be equivalent to those of \cite{Donos:2006ms}, but
we have not verified this.

\subsection{$AdS_3$ from K\"ahler 2 in $CY_2$}

\paragraph{AdS radial direction from overall transverse space}
For this case it is possible for the radial direction to come from the
overall transverse space. One finds that $\lambda$ must be a constant, which we take to be 1,
and that the solution is the well known $AdS_3\times S^3\times CY_2$ solution:
\bea
\dd s^2&=&\dd s^2(AdS_3)+\dd s^2(S^3)+\dd s^2(\mathcal{M}_4),\nn
\Theta&=&\mbox{Vol}[AdS_3]\w J_4.
\eea

\paragraph{AdS radial direction from frame rotation}
Alternatively, carrying out the general frame rotation and writing
$\dd s^2({\cal M}_4)=(e^1)^2+(e^2)^2+ (e^3)^2 +(\hat u)^2$
we instead find that the metric and flux are
\bea
\dd s^2&=&\frac{1}{\l }\left[\dd
  s^2(AdS_3)+\lambda^2\left(\frac{1}{1-\l^2\rho^2}\dd\r^2+\r^2\dd
    s^2(S^3)\right)\right]\nn&&+e^1\otimes e^1+e^2\otimes e^2+e^3\otimes e^3,\nn
\Theta&=&\mbox{Vol}[AdS_3]\w[\dd(\l^{-3/2}\sss
e^3)-2(\l^{-1}e^{12}+\r e^3\w\hat{\r})].
\eea
The torsion conditions are that the $S^3$
directions are Killing, $\dd \l$ has no $e^3$ component, together with 
\bea
e^{1,2}&=&\frac{\l^{1/2}}{\sss}\dd x_{1,2},\nn
\dd
\left(\frac{\l^{1/2}}{\sss}e^3\right)&=&
-\frac{1}{\rho}*_3\dd\left(\frac{1}{1-\l^2\rho^2}\right),
\eea
where $*_3$ denotes the Hodge dual on the three-manifold with
metric and orientation given by
\bea
\dd s^2&=&\dd x^2_1+\dd x_2^2+\dd\r^2,\nn
\mbox{Vol}&=&\dd x_1\w\dd x_2\w\dd\r.
\eea
along with $\dd*_{10}\Theta=0$. 
These conditions imply that we can
introduce a coordinate $\psi$ such 
$e^3=\l^{-1/2}\sss(\dd\psi+B)$ with $\partial_\psi$ a Killing vector.
These conditions are equivalent to
those of LLM \cite{Lin:2004nb}. The Bianchi identity is implied by the
torsion conditions; the ten-dimensional Hodge dual of $\Theta$ 
is given by
\bea
{*}_{10}\Theta=-\mbox{Vol}[S^3]\w\Big[\dd(\l^{1/2}\rho^2\sss
e^3)+2\left(\l\r^2e^{12}+\r e^3\w\hat{\r}\right)\Big].
\eea

These geometries are dual to two-dimensional SCFTs with $(4,4)$ supersymmetry
and the $SO(4)\times U(1)$ symmetry of the solution is dual to the $R$-symmetry.
We are unaware of any explicit $AdS$ examples in this class. 
After analytic continuation we recover the 1/2 BPS 
LLM bubbling solutions with $\bbR\times SO(4)\times SO(4)\times U(1)$ symmetry.

\subsection{$AdS_3$ from K\"{a}hler-2 in $CY_4$}
This case is different from the previous cases in that
the wrapped-brane spacetime does not have any overall
transverse directions. It was first derived from a 
wrapped-brane geometry in \cite{jn}. In the notation of this paper, if we 
write\footnote{Note that this means that the conformally rescaled metric
$L\dd s^2({\cal M}_8)$ is a cone: 
\be
L\dd s^2({\cal M}_8)=e^{2r}\dd r^2+e^{2r}\lambda\left(
\dd s^2(\mathcal{M}_6)+e^7\otimes e^7\nn\right)
\ee
and so this case is rather analogous to the Sasaki-Einstein case.}
 $L=\l e^{2r}$, $e^8=\l^{-1/2}\dd r$, then equations \eqref{mcy41}
and \eqref{mcy42} lead to a metric and flux given by
\bea
\dd s^2&=&\frac{1}{\l}\dd s^2(AdS_3)+
\dd s^2(\mathcal{M}_6)+e^7\otimes e^7,\nn
\Theta&=&\mbox{Vol}[AdS_3]\w[\dd(\l^{-3/2}e^7)-2J_6],
\eea
with the $SU(3)$ structure satisfying
\bea
\dd(\l^{-1}J_6)&=&0,\nn
J_6^2\w\dd(\l^{1/2}e^7)&=&\frac{2}{3}\l J_6^3,\nn
\dd(\l^{-3/2}\Omega_6)&=&2i\l^{-1}e^7\w\Omega_6.
\eea
Observe that in this case not all the torsion modules are fixed by the 
above conditions. As a result, in this case it is necessary to impose
$\dd *_{10}\Theta=0$ as an extra condition.
One can show that these are equivalent to the conditions of \cite{Kim:2005ez}.
In particular we note that these conditions imply that we can
introduce a coordinate $\psi$ such 
$e^7=\l^{-1/2}(\dd\psi+B)$ with $\partial_\psi$ a Killing vector
that preserves $J_6$ and $\lambda$ but not $\Omega_6$. Also the
metric $\lambda^{-1}\dd s^2({\cal M}_6)$ is K\"ahler. 

These geometries are dual to two-dimensional SCFTs with $(0,2)$ supersymmetry and the
$U(1)$ Killing vector is dual to the $R$-symmetry .
A rich set of examples of solutions of these equations can be found in
\cite{naka,Gauntlett:2006af,Gauntlett:2006qw,Gauntlett:2006ns}.

\section{Explicit examples}
In this section we will study two explicit solutions in detail -
the $AdS_3$ solution of \cite{mn} that is dual to a SCFT with $\mathcal{N}=(2,2)$ 
supersymmetry and the $AdS_2$ solution of \cite{oz} dual to a SCQM with 2 supercharges. 
These arise from $D3$-branes wrapping K\"{a}hler two-cycles in
CY three-folds and associative three-cycles, respectively. In each case,
the first part of our investigation will be to verify that these solutions
satisfy our $AdS$ conditions, by making their $G$-structure manifest. This
serves as a rigid consistency check of our conditions.  

In the second part of our investigation we will
frame-rotate back to the canonical Minkowski frame.
In other words we will write the solutions in a way in which the wrapped-brane $G$-structure
of section 2, defined by half of the Killing spinors, is manifest.
Inspired by the form of the metrics when re-written
in this fashion, we can construct an ansatz for a more general class of wrapped-brane geometries
that could describe an interpolation from a special holonomy metric to the $AdS$ fixed
point. Given that the $AdS$ solutions describe the near horizon limit of
$D3$-branes wrapping calibrated cycles, such interpolating solutions should exist. 
We show that our ansatz does indeed include singular special holonomy metrics that have a
calibrated cycle of the form that appears in the $AdS$ solutions.
The partial differential equations that need to be solved in order to construct interpolating solutions
are involved and we have not managed to find any solutions.

\subsection{$AdS_3$ from K\"{a}hler-2 in $CY_3$}
An explicit solution of this type was first constructed in \cite{mn}.
The solution was first constructed in gauged supergravity and then uplifted to type IIB.
It describes the near horizon limit of a $D3$-brane wrapping a holomorphic $H^2$ in a $CY_3$.
The $H^2$ can also be replaced with a discrete quotient, $H^2/\Gamma$ and hence a compact Riemann surface with genus $g> 1$.
The metric can be written
\bea\label{nnn}
\dd s^2&=&\frac{1}{\l }\Big[\dd s^2(AdS_3)+\dd
  s^2(H^2)+\frac{\l^2}{1-\l^2\r^2}\dd \r^2+\l^2\r^2\dd
    s^2(S^1)\nn
&&+(1-\l^2\r^2)\Big(\dd
    s^2(S^2)+(\dd\psi+P-P')^2\Big)\Big],
\eea
where
\bea
\l^2&=&\frac{8}{1+4\rho^2},\nn
\dd P&=&\mbox{Vol}(S^2),\nn
\dd P'&=&\mbox{Vol}(H^2).
\eea
with $\rho\in[0,1/2]$. 
Observe that the trivial $S^1$ fibre can be taken to smoothly degenerate at $\rho=0$, 
while an $S^3$ smoothly degenerates at $\rho=1/2$ if $\psi$ has period $4\pi$. 

We define the frame
\bea
e^1+ie^2&=&\frac{1}{\l^{1/2}}e^{i\psi/2}(\dd \mu+i\sinh\mu\dd\beta),\nn
e^3+ie^4&=&\frac{\sss}{\l^{1/2}}e^{i\psi/2}(\dd \theta+i\sin\theta\dd\phi),\nn
e^5&=&\frac{\sss}{\l^{1/2}}(\dd\psi-\cos\theta\dd\phi-\cosh\mu\dd\beta),
\eea
where $\mu,\beta$ are coordinates for $H^2$ and $\theta,\phi$ are
coordinates for $S^2$. We can then define an $SU(2)$ structure by
\bea
J_4&=&e^{12}+e^{34},\nn
\Omega_4&=&(e^1+ie^2)\w (e^3+ie^4).
\eea
It may be verified explicitly that this is a solution of the
conditions that were presented in section 3.3. 

We now introduce an ansatz that could describe an interpolating solution from
a $CY_3$ metric to the above $AdS_3$ solution. 
To do this it is illuminating to first 
identify the $SU(3)$ structure of the $AdS_3$ solution, regarded
as a solution with a two-dimensional Minkowski factor. We therefore
rotate back from the $AdS$ to the Minkowski frame, using
the formulae of section 3. We find that the basis one-form $e^6$ of the Minkowski
frame is given by
\bea
e^6&=&L^{1/2}e^{-r/2}\dd\left(-2e^{-r/2}\sqrt{\frac{1-4\r^2}{8}}\right).
\eea
Defining the Minkowski-frame coordinate
\bea
u=-2e^{-r/2}\sqrt{\frac{1-4\r^2}{8}},
\eea
the metric may be written as
\bea\label{mnb}
\dd s^2&=&L^{-1}\Big[\dd s^2(\bbR^{1,1})+F^2\dd
s^2(H^2)\Big]\nn&+&L\Big[F^{-1}\Big(\dd u^2+\frac{u^2}{4}\Big[\dd
s^2(S^2)+(\dd\psi+P-P')^2\Big]\Big)+\dd t^2+t^2\dd s^2(S^1)\Big],
\eea
where 
\bea
F=e^{r}=-\frac{u^2}{4z^2}+\frac{1}{4z^2}\sqrt{u^4+4z^2}.
\eea
We also find that $L=2F^2/\sqrt{1-u^2F}$.
The $SU(3)$ structure is given by the standard form
\bea
J_6&=&e^{12}+e^{34}+e^{56},\nn
\Omega_6&=&(e^1+ie^2)\w (e^3+ie^4)\w (e^5+ie^6),
\eea
with the Minkowski frame following from the $AdS$ frame given by
\bea\label{mink}
e^1+ie^2&=&L^{-1/2}Fe^{i\alpha\psi}(\dd \mu+i\sinh\mu\dd\beta),\nn
e^3+ie^4&=&-\frac{L^{1/2}u}{2F^{1/2}}e^{i\gamma\psi}(\dd \theta+i\sin\theta\dd\phi),\nn
e^5&=&-\frac{L^{1/2}u}{2F^{1/2}}(\dd\psi+P-P'),\nn
e^6&=&L^{1/2}F^{-1/2}\dd u.
\eea
By construction this structure satisfies \reef{mcy31}, \reef{mcy32}, and the Bianchi identity.

We can now make the following ansatz for the interpolating solution:
\bea\label{mnbv}
\dd s^2&=&L^{-1}\Big[\dd s^2(\mathbb{R}^{1,1})+F_1F_2\dd s^2(H^2)\Big]\nn
&+&L\Big[F_1^{-1}\Big(\dd u^2+\frac{u^2}{4}(\dd \psi
+P-P')^2\Big)+F_2^{-1}\frac{u^2}{4}\dd s^2(S^2)+\dd t^2+t^2 ds^2
(S^1)\Big],\nn&&
\eea
with $L$ and $F_{1,2}$ arbitrary functions of $u,z$. We impose as a
boundary condition that this metric smoothly matches on to \eqref{mnb}
in the $AdS$ limit. Now we wish to determine the other boundary
condition, by finding the most general special holonomy metric of the
form \eqref{mnbv}. For special holonomy, we must have that $L=1$, that
$F_{1,2}$ are functions of $u$ only, and that $J_6$, $\Omega_6$, with the
obvious frame, are closed.  It is easy to verify that $\Omega_6$ is closed
for any choice of $F_1,F_2$.
Closure of $J_6$ then implies the equations
\bea
\partial_u(F_1F_2)+\frac{u}{2F_1}&=&0,\nn
\partial_u\left(\frac{u^2}{4F_2}\right)-\frac{u}{2F_1}&=&0.
\eea
Adding and integrating, we find
\bea\label{okl}
F_2=\frac{a^2\pm\sqrt{a^4-F_1u^2}}{2F_1},
\eea
for some positive constant $a^2$. Inserting this into one of the
remaining equations, making the substitution
$F_1=a^4u^{-2}\cos^2\xi$ and integrating, we get
\bea\label{opl}
-\frac{1}{3}\sin^3\xi+\sin\xi=b\mp\frac{u^4}{4a^6}.
\eea
By sending $\xi\rightarrow-\xi$, $b\rightarrow-b$, we can choose the
upper sign in \eqref{okl} and \eqref{opl}. 
Furthermore in order to obtain a smooth degeneration when $u=0$ 
we will choose $b=2/3$. A final change of coordinates
\be
\sin\xi=1-\frac{r^2}{3a^2}
\ee
allows us to cast the metric in the form
\be\label{po2}
ds^2=\frac{6a^2-r^2}{6}\dd s^2(H^2)
+\frac{r^2}{6}\dd s^2(S^2)
+\kappa^{-1} \dd r^2+\frac{\kappa r^2}{9}(\dd \psi+P-P')^2
\ee
where
\be
\kappa=\frac{9a^2-r^2}{6a^2-r^2}
\ee
In this form it is clear that the metric is a hyperbolic analogue of the
well known metric on the resolved conifold constructed in \cite{Candelas:1989js,Pando Zayas:2000sq}, which has
a holomorphic $S^2$. Note that at $r=0$, provided that $\psi$ has period $4\pi$ (as in
the $AdS_3$ solution), an $S^3$ smoothly degenerates leaving a holomorphic $H^2$.
At $r^2=6a^2$, however, the metric is singular (unlike the resolved conifold metric).
It is natural to view \reef{po2} as a good local model
of a holomorphic $H^2$, for which we can consider wrapping $D3$-branes. The gauged supergravity
solution then describes the smooth back-reacted geometry in the near horizon limit\footnote{Following
\cite{Mac Conamhna:2007ag} we might also try to interpret the gauged supergravity solution as describing the back reacted
geometry of $D3$-branes wrapping a singular holomorphic two-cycle at
$r^2=6a^2$.}. 

Upon substituting the ansatz \reef{mnbv} into the conditions for a wrapped
brane geometry we obtain some complicated p.d.e.'s for $L$, $F_1$ and $F_2$
which we won't write down. We conjecture that they admit a solution interpolating
between the above special holonomy metric and the $AdS$ solution. 

\subsection{$AdS_2$ from associative}
An explicit solution of this type was constructed in five dimensional
gauged supergravity in \cite{oz} and then uplifted to IIB supergravity. 
It describes the near-horizon limit of a $D3$-brane wrapping an 
associative hyperbolic 3-space, $H^3$.
We can also replace $H^3$ with a compact discrete 
quotient $H^3/\Gamma$ without breaking supersymmetry.
Correcting the expression of \cite{oz} upon lifting 
to ten dimensions\footnote{Our expression differs from \cite{oz} by a factor 2 in the radius of the five-sphere.} , 
we find that the metric of the IIB solution is given by
\bea
\dd s^2&=&\frac{1}{\l }\Big[\dd s^2(AdS_2)+4\dd
  s^2(H^3)+\frac{\l^2}{1-\l^2\r^2}\dd\rho^2+\l^2\r^2\dd
    s^2(S^1)\nn&&+4(1-\l^2\r^2)\tilde{\mu}^a\tilde{\mu}^a\Big],
\eea
Here $\dd s^2(H^3)$ is the maximally symmetric metric on $H^3$ (with Ricci scalar equal to -6).
If we introduce left-invariant one-forms $\s^a$ on $S^3$ satisfying
$\dd\s^a=\frac{1}{2}\e^{abc}\s^b\w\s^c$, and the spin connection
$\omega_{ab}$ for $\dd s^2(H^3)$, then
\bea
\tilde{\mu}^a=\s^a-\frac{1}{2}\e^{abc}\omega_{bc}.
\eea
In addition
\bea
\l^2&=&\frac{64}{1+48\rho^2},
\eea
and we take $\rho\in [0,1/4]$. 
The $S^1$ smoothly
degenerates at $\rho=0$, while at $\rho=1/4$ the $S^3$ smoothly
degenerates. Defining the frame
\bea
e^a&=&\frac{2}{\l^{1/2}}\tilde{e}^a,\nn
\mu^a&=&\frac{2}{\l^{1/2}}\sss\tilde{\mu}^a,
\eea
where $\tilde{e}^a$ is a basis for $\dd s^2(H^3)$ (and hence $d\tilde{e}^a+\omega^a{}_b \tilde{e}^b=0$),
the $SU(3)$ structure of this solution is given by
\bea
J_6&=&\mu^a\w e^a,\nn
\Omega_6&=&\frac{1}{6}\e^{abc}(\mu^a+ie^a)\w(\mu^b+ie^b)\w(\mu^c+ie^c).
\eea
Using the equations (9.64)-(9.69) of \cite{Gauntlett:2006ux}, it is easy to verify
that this is an exact solution of the torsion conditions and Bianchi
identity that we derived in section 3.1.

We now discuss an ansatz that could describe an interpolation from a $G_2$ holonomy metric 
to this $AdS_2$ solution. As in the previous subsection, we
first obtain the Minkowski $G_2$ structure for the 
$AdS_2$ solution. We find that the one-form $e^7$ in the Minkowski 
frame is given by
\bea\label{e7}
e^7=L^{1/2}e^{-3r/4}\dd\left(-\frac{1}{2}e^{-r/4}\sqrt{1-16\rho^2}\right).
\eea
Defining the Minkowski frame coordinate
\bea
u=-\frac{1}{2}e^{-r/4}\sqrt{1-16\rho^2},
\eea
the metric of the $AdS_2$ solution can be written
\bea\label{g2ads}
\dd s^2=L^{-1}\Big[-\dd T^2+4F\dd s^2(H^3)\Big]+L\Big[F^{-3/4}\Big(\dd
u^2+\frac{u^2}{4}\tilde{\mu}^a\tilde{\mu}^a\Big)+\dd z^2+z^2\dd\phi^2\Big],
\eea
where
\bea
F=e^{2r},
\eea
and $e^r$ is given in terms of $u$ and $z$ by a (positive-signature
metric inducing) root of the
quartic 
\bea
256z^4e^{4r}-32z^2e^{2r}-16u^4e^r+1=0.
\eea
We also find  that $L=4F/(1-3u^2F^{1/4})^{1/2}$.
By construction, the $G_2$ structure given by 
\bea
\varphi&=&J\w e^7-\mbox{Im}\Omega,\nn
{*}_7\varphi&=&\frac{1}{2}J\w J+\mbox{Re}\Omega\w e^7,
\eea
with $e^7$ given by \eqref{e7}, satisfies the
the equations \reef{awbc}, including the Bianchi identity. 

Based on this result, we can now make the following ansatz for an interpolating solution:
\bea\label{inter}
\dd s^2=L^{-1}\Big[-\dd T^2+F_1^2\dd s^2(H^3)\Big]+L\Big[F_3^3\dd
u^2+F_2^2\tilde{\mu}^a\tilde{\mu}^a+\dd z^2+z^2 \dd\phi^2\Big],
\eea
with $L$, $F_{1,2,3}$ arbitrary functions of $u$ and $z$, and a
boundary condition on the interpolating solution such that \eqref{inter}
smoothly matches on to \eqref{g2ads} in the $AdS$ limit. The second
boundary condition on the interpolating solution is that it should
smoothly match onto a $G_2$ metric, which we now determine. Requiring
that \eqref{inter} is a metric of $G_2$ holonomy implies that $L$
is a constant, which we set to 1, that $F_{1,2,3}$ are functions of
$u$ only, and that $\varphi$ and $*_7\varphi$, with the obvious
frame, are closed. Since $F_3$ is a function of $u$ alone we can choose it 
to be 1. Then imposing closure of $\varphi$ we
find the conditions 
\bea\label{closph}
\frac{1}{3}\partial_uF_1^3-F_1F_2&=&0,\nn
\partial_u(F_1F_2^2)+F_1F_2&=&0.
\eea
Closure of $*_7\varphi$ produces the condition
\bea
\partial_u(F_1^2F_2^2)=F_2^3-F_1^2F_2,
\eea
which is implied by \eqref{closph}. It is straightforward to integrate
\eqref{closph}; adding, we immediately obtain
\bea
F_2=\sqrt{\frac{\a}{F_1}-\frac{F_1^2}{3}},
\eea
for some constant $\a$. Then we have
\bea
\partial_uF_1=-\sqrt{\frac{\a}{F_1^2}-\frac{1}{3}}.
\eea
Defining a new coordinate $x$ according to 
\bea
\partial_u=\sqrt{\frac{\a}{F_1^2}-\frac{1}{3}}\partial_x,
\eea
we get 
\bea
F_1=x+\b.
\eea 
The constant $\b$ may be eliminated by a shift in $x$. Finally,
defining $x=(3\a)^{1/3}R$, then dropping the tildes together with
an overall scale factor of $3(3\a)^{2/3}$, the $G_2$ metric is
\bea\label{po}
\dd s^2=\frac{\dd R^2 }{\frac{1}{R^3}-1}+\frac{R^2}{3}\dd
s^2(H^3)+\frac{R^2}{9}\Big(\frac{1}{R^3}-1\Big)\tilde{\mu}^a\tilde{\mu}^a.
\eea 
Observe that this metric is the hyperbolic analogue of the well-known $G_2$ metric on an
$R^4$ bundle over $S^3$, \cite{bryant}, \cite{pope}, which has an associative
$S^3$. In particular, at $R=1$ the metric is smooth and describes an associative $H^3$.
Unlike the metric in \cite{bryant}, \cite{pope}, however, the metric \reef{po} is
singular at $R=0$. It is natural to view \reef{po} as a good local model
of an associative $H^3$, for which we can consider wrapping $D3$-branes. The gauged supergravity
solution then describes the smooth back-reacted geometry in the near horizon limit\footnote{Following
\cite{Mac Conamhna:2007ag} we might also try to interpret the gauged supergravity solution as describing the back reacted
geometry of $D3$-branes wrapping a singular associative three-cycle at $R=0$.}.

Upon substituting the ansatz \reef{inter} into the conditions for a wrapped
brane geometry we obtain some complicated p.d.e.'s for $L$, $F_1$, $F_2$ and $F_3$
which we won't write down. We conjecture that they admit a solution interpolating
between the above special holonomy metric and the $AdS$ solution.

\section{Conclusions}
In this paper we have derived an interesting class of supersymmetric geometries of
type IIB supergravity with Minkowski factors and five-form flux
that are associated with $D3$-branes wrapping calibrated cycles in special holonomy manifolds
(or configurations of intersecting $D3$-branes). 
Using these wrapped-brane geometries we determined the extra conditions that are
required in order to obtain a supersymmetric solution with an $AdS$ factor.
The $AdS_2$ conditions we have derived for the cases of $D3$-branes wrapping associative or SLAG three-cycles, are new. 
Although the $AdS_3$ conditions for the cases of $D3$-branes wrapping holomorphic two-cycles in $CY_2$, $CY_3$ and
$CY_4$ have been derived before, here we make an explicit link between these
geometries and the wrapped-brane geometries.

By analytic continuation of the $AdS$ metrics and torsion conditions,
one obtains the conditions defining a class of supersymmetric 
geometries containing spheres. The class of BPS geometries with the $AdS_3$ replaced by $S^3$
have been classified before, but we found a new class with an $S^2$ factor by replacing $AdS_2$
with $S^2$ for the case associated with associative three-cycles. These geometries preserve 1/8 supersymmetry
and have $\bbR\times SU(2)$ symmetry. It would be interesting to study them further.

For two explicit $AdS$ solutions, we have
verified that they satisfy the appropriate torsion conditions, by
explicitly obtaining their structures. This serves as a strong overall
consistency check of our results. For these solutions we also
constructed a more general ansatz for the corresponding wrapped-brane geometries
which could describe solutions that interpolate between a special holonomy
metric and the $AdS$ solution. In particular, we show that the ansatz admits
singular special holonomy metrics that have calibrated cycles 
of the appropriate type. It would be very interesting 
to construct explicit interpolating solutions and to study how the 
singularity of the special holonomy metric gets resolved in the $AdS$ limit.

\subsection*{Acknowledgements}
We would like to thank Vijay Balasubramanian, Jaume Gomis, Jan Gutowski, Nakwoo Kim,
Joan Simon, Nemani Suryanarayana and Daniel Waldram
for helpful discussions. JPG is supported by an EPSRC Senior Fellowship and a Royal Society
Wolfson Award. OC is supported by an EPSRC Research Fellowship. JPG would like to thank the Galileo Galilei Institute for Theoretical
Physics for hospitality.

\end{document}